\documentclass{jfm}

\usepackage{color}

\begin{document}

\newcommand{\arcsec}{\ifmmode^{\prime\prime}\else $^{\prime\prime}$\fi}
\newcommand{\arcmin}{\ifmmode^{\prime}\else $^{\prime}$\fi}
\newcommand{\degrees}{\ifmmode^{\circ}\else $^{\circ}$\fi}
\newcommand{\kms}{km\,s$^{-1}$}
\newcommand{\ms}{m\,s$^{-1}$}
\newcommand{\micron}{$\mu$m}
\newcommand{\smallhalf}{\textstyle \frac{1}{2} \displaystyle}
\newcommand{\smallonethird}{\textstyle \frac{1}{3} \displaystyle}
\newcommand{\smallonefourth}{\textstyle \frac{1}{4} \displaystyle}
\newcommand{\smalltwothirds}{\textstyle \frac{2}{3} \displaystyle}
\newcommand{\smallfourthirds}{\textstyle \frac{4}{3} \displaystyle}
\newcommand{\smallonequarter}{\textstyle \frac{1}{4} \displaystyle}
\newcommand{\bfx}{{\bf x}}
\newcommand{\bfk}{{\bf k}}
\newcommand{\bfK}{{\bf K}}
\newcommand{\bfc}{{\bf c}}
\newcommand{\bfe}{{\bf e}}
\newcommand{\bff}{{\bf f}}
\newcommand{\bfg}{{\bf g}}
\newcommand{\bfq}{{\bf q}}
\newcommand{\bfr}{{\bf r}}
\newcommand{\bfu}{{\bf u}}
\newcommand{\bfU}{{\bf U}}
\newcommand{\bfv}{{\bf v}}
\newcommand{\bfw}{{\bf w}}
\newcommand{\bfb}{{\bf b}}
\newcommand{\bfB}{{\bf B}}
\newcommand{\bfH}{{\bf H}}
\newcommand{\bfI}{{\bf I}}
\newcommand{\bfE}{{\bf E}}
\newcommand{\bfD}{{\bf D}}
\newcommand{\bfj}{{\bf j}}
\newcommand{\bfa}{{\bf a}}
\newcommand{\bfn}{{\bf n}}
\newcommand{\bfs}{{\bf s}}
\newcommand{\bfA}{{\bf A}}
\newcommand{\bfF}{{\bf F}}
\newcommand{\bfM}{{\bf M}}
\newcommand{\bfP}{{\bf P}}
\newcommand{\bfQ}{{\bf Q}}
\newcommand{\bfS}{{\bf S}}
\newcommand{\bfT}{{\bf T}}

\newcommand{\bfgamma}{{\mbox{\boldmath $\gamma$}}}
\newcommand{\bfkappa}{{\mbox{\boldmath $\kappa$}}}
\newcommand{\bfOmega}{{\mbox{\boldmath $\Omega$}}}
\newcommand{\bfomega}{{\mbox{\boldmath $\omega$}}}
\newcommand{\bfsigma}{{\mbox{\boldmath $\sigma$}}}
\newcommand{\bftau}{{\mbox{\boldmath $\tau$}}}
\newcommand{\bfxi}{{\mbox{\boldmath $\xi$}}}

\newcommand{\bfnu}{{\mbox{\boldmath $\nu$}}}

\newcommand{\bfnabla}{{\mbox{\boldmath $\nabla$}}}

\newcommand{\bfitl}{{\mbox{\boldmath $l$}}}

\newcommand{\calC}{{\mathcal C}}
\newcommand{\calS}{{\mathcal S}}
\newcommand{\calV}{{\mathcal V}}
\newcommand{\calE}{{\mathcal E}}
\newcommand{\calP}{{\mathcal P}}
\newcommand{\calR}{{\mathcal R}}
\newcommand{\calT}{{\mathcal T}}
\newcommand{\calH}{{\mathcal H}}
\newcommand{\calO}{{\mathcal O}}

\newtheorem{lemma}{Lemma}
\newtheorem{corollary}{Corollary}

\shorttitle{Peristaltic pumping in annular tubes} %for header on odd pages
\shortauthor{J. B. Carr, J. H. Thomas, J. Liu, J. K. Shang} %for header on even pages

\title{Peristaltic pumping in thin, non-axisymmetric, annular tubes}

\author
 {
 J. Brennen Carr\aff{1}
  \corresp{\email{jcarr12@ur.rochester.edu}},
  John H. Thomas\aff{1},
  Jia Liu\aff{1},
  \and
  Jessica K.  Shang\aff{1}
  }

\affiliation
{\aff{1}
Department of Mechanical Engineering, University of Rochester, Rochester, NY 14627, USA}

\maketitle

\begin{abstract}

Two-dimensional laminar flow of a viscous fluid induced by peristalsis due to a moving wall wave has been studied previously for a rectangular channel, a circular tube, and a concentric circular annulus. 
Here we study peristaltic flow in a non-axisymmetric annular tube, where the flow is three-dimensional, with azimuthal motions. 
This geometry is motivated by experimental observations of cerebrospinal fluid flow along perivascular spaces (PVSs) surrounding arteries in the brain, which is at least partially driven by peristaltic pumping. 
These PVSs are well matched, in cross-section, by an adjustable model consisting of an inner circle (arterial wall) and an outer ellipse (outer edge of the PVS), not necessarily concentric. 
We use this model, which may have other applications, as a basis for numerical simulations of peristaltic flow. 
We use a finite-element scheme to compute the flow driven by a propagating sinusoidal radial displacement of the inner wall. 
Unlike peristaltic flow in a concentric circular annulus, the flow is fully three-dimensional, with streamlines wiggling in both the radial and axial directions. 
We examine the dependence of the flow on the elongation of the outer elliptical wall and on the eccentricity of the configuration. 
We find that time-averaged volumetric flow decreases with increasing ellipticity or eccentricity. 
Azimuthal pressure variations, caused by the wall wave, drive an oscillatory azimuthal flow in and out of the narrower gaps. 
The additional shearing motion in the azimuthal direction will enhance Taylor dispersion in these flows, an effect that might have practical applications.

\medskip

{\bf Key words:} Peristaltic pumping, cerebrospinal fluid
\end{abstract}

\section{Introduction}

Peristaltic pumping occurs when a wave of area contraction and expansion propagates along the length of a 
flexible channel or tube filled with a liquid \citep{jaffrin1971}. This mechanism is the basis for roller and finger pumps,
used in cases where it is necessary to keep the transported liquid from coming into contact with the pump itself.
The mechanism also occurs in the perivascular spaces (PVS) surrounding arteries in the brain, in which cerebrospinal fluid
is pumped by radial pulsations of the artery wall induced by the heartbeat  \citep{mestre2018, bedussi2017}.

Various mechanisms have been proposed as a driving force for cerebrospinal flow in the brain, including an overall pressure gradient due to the
production of the fluid, a varying pressure gradient due to respiration, and peristaltic pumping by arterial wall motions. 
Recent {\it in vivo} experiments with mice show that there is indeed pulsatile flow in the perivascular spaces around arteries in the brain \citep{bedussi2017, mestre2018}.
The particle-tracking experiments of \citet{mestre2018} show that the flow is driven mainly by peristaltic pumping due to motions of
the arterial wall driven by the cardiac cycle, with net (time-averaged) flow in the same direction as the blood flow, a mechanism aptly
named ``perivascular pumping'' by \citet{Hadaczek2006}. \citet{mestre2018} provide detailed measurements of the velocity field of the flow, the motions of the arterial wall, and the actual size and shape of the perivascular spaces around the arteries. They also show that the perivascular spaces around 
arteries are essentially open (unobstructed), not porous (see also \citet{minrivas2020}).

\begin{figure}
  \centerline{\includegraphics[height=4cm]{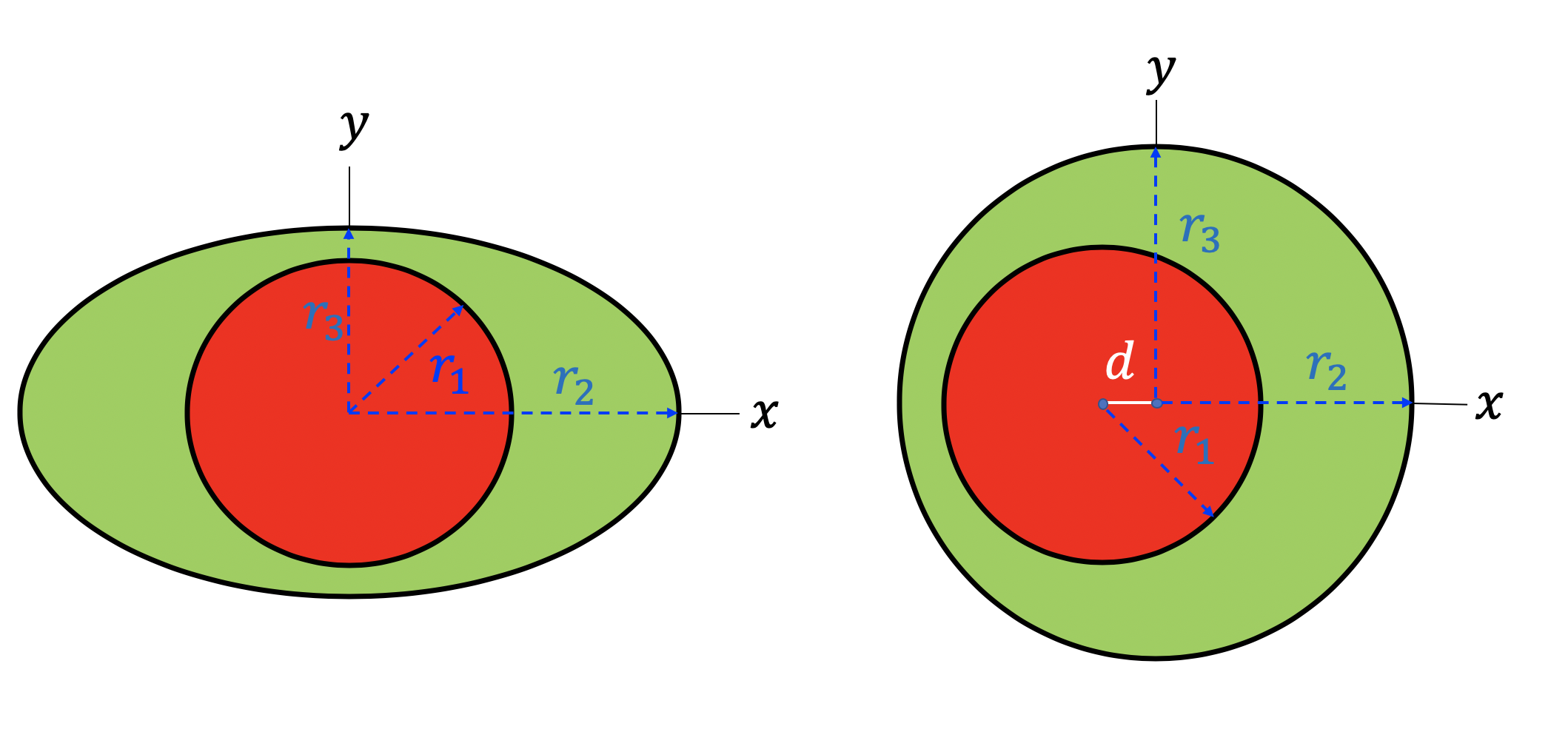}}
  \caption{Cross-sections of the annular tube for the concentric elliptical annulus model (left) 
  and the eccentric circular annulus model (right).
  }
 \label{fig:model}
\end{figure}

\citet{Tithof2019} have shown that the various shapes of the cross-sections of PVSs around mouse surface and penetrating
arteries, observed {\it in vivo},  can be fit quite well with a simple geometric model, consisting of a circular inner boundary (the artery)
and an elliptical outer boundary (the outer wall of the PVS), allowing the circle to be eccentric with respect to the ellipse. 
We adopt the same geometric model in our present study of peristaltic pumping (see fig. \ref{fig:model}). 
\citet{Tithof2019} compute the hydraulic resistance for annular channels with this model cross-section, varying the
ellipticity and eccentricity, and show that the actual shapes of PVSs are near optimal, in the sense of offering the
least hydraulic resistance for a given cross-sectional area. Here, we also compute hydraulic resistances (for steady Poiseuille flow)
and compare them with the results of \cite{Tithof2019} in order to validate our finite-element code.

There have been a few attempts to model peristaltic pumping in perivascular spaces \citep{bilston2003, schley2006, wang2011, coloma2016}: see  the review by \citet{thomas2019}. These studies model the perivascular space as either a two-dimensional channel or a concentric circular annulus. The most relevant study for our 
purposes is that of \citet{wang2011}: they model the perivascular space as a concentric circular annulus, filled with a porous medium (instead of an open channel,
as we assume in this paper), with flow driven by a sinusoidal wall wave propagating along the inner boundary. By making the long wavelength,
low Reynolds number approximation \citep{shapiro1969, jaffrin1971}, they are able to obtain an analytical solution. Here we
present (in Appendix A) an analytical solution to the same problem, except with an open rather than porous annulus and use it as a check on our numerical simulations.

The analytical solution demonstrates that the effectiveness of the pumping scales as $(b/\ell)^2$, where $b$ is the amplitude of the wall wave and $\ell$ is the width of the gap 
between the inner and outer boundaries. 
We can use this result as a basis for imagining what the peristaltic flow is like, qualitatively, in our elliptical, non-concentric model. 
For a concentric circular annulus and an axisymmetric wall wave, the flow is axisymmetric and the instantaneous streamlines lie in planes 
through the central axis. With an elliptical outer boundary, or an eccentric circular annulus, we can expect the flow to be 
three-dimensional, with secondary motions in the azimuthal direction. As in the circular annulus, the effectiveness 
of the pumping scales as $(b/\ell)^2$, where $b$ is the amplitude of the wall wave and $\ell$ is the width of the gap between the 
inner and outer boundaries. For a concentric circular annulus, the gap width $\ell$ and hence the 
pumping effectiveness are axisymmetric, so the resulting flow is also axisymmetric and the streamlines lie
in planes through the central axis and wiggle only in the radial direction. For an elliptical outer boundary, or an eccentric circular annulus,  
the gap width $\ell$, and hence the pumping effectiveness, vary in the azimuthal direction, so there will be pressure variations 
in the azimuthal direction (with higher pressure where the gap is narrower) that  drive a secondary, oscillatory flow in the azimuthal direction, causing
streamlines to also wiggle in the azimuthal direction.

To our knowledge, this is the first study of peristaltic pumping in elliptical and eccentric annuli: hence, our results may be of 
interest in contexts other than perivascular pumping. 
   
\section{Computational methodology}

\subsection{The computational model and its scaling}

We consider the flow of a Newtonian viscous fluid in an annular tube, driven by a radial displacement of the inner wall 
in the form of a sinusoidal wave propagating in the axial direction.  The annular space between the inner and outer walls 
of the tube is assumed to be open (i.e., unobstructed), not porous.  The annular tube is uniform in
the axial direction. Its cross-section consists of a circular inner wall and an elliptical outer wall, and we allow the circle 
to be eccentric with respect to the ellipse (see fig. \ref{fig:model}). 
The inner wall is deformable, and peristaltic pumping is driven by a radial displacement of the form (in cylindrical coordinates)
	\begin{equation}
	r(z,t) = r_1 + b\, sin\bigg(\frac{2 \pi}{\lambda} (z - c\,t)\bigg) ,
	\label{eq:wallwave}
	\end{equation}
where $b$ is the wave amplitude, $c$ is the wave speed, and $\lambda$ is the wavelength. 

We nondimensionalize the geometric quantities using scalings similar to those of \citet{Tithof2019}. The radii $r_2$ and $r_3$ of the ellipse, 
the eccentricity $d$ (for the circular annulus), and the wave amplitude $b$ are scaled as
	\begin{equation}
	\alpha = \frac{r_2}{r_1}, \quad \beta = \frac{r_3}{r_1}, \quad \epsilon  = \frac{d}{r_2}, \quad b^* = \frac{b}{r_1} .
	\label{eq:scaling1}
	\end{equation}
For the elliptical annuli, it is useful to use $\alpha / \beta$ as a measure of the elongation of the outer ellipse.

In our simulations, the ratio of the area $A_{\rm pvs}$ of the annular region enclosed by the two boundaries (the perivascular space) and the area $A_{\rm art}$ of the inner circle
(the artery) was held constant at 1.4, an average value measured {\it in vivo} \citep{mestre2018}. Then $\alpha$ and $\beta$ are related, allowing all models to be defined 
by a single parameter. The flow velocity $\bfv = (v_r, v_{\theta}, v_z)$,  volume flow rate $Q$, and pressure are nondimensionalized using the wave speed $c$,
as follows:
	\begin{equation}
	\bfv^* = \frac{\bfv}{c}, \quad Q^* = \frac{Q}{c A_{\rm pvs}}, \quad p^* = \frac{\lambda}{\mu c} p ,
	\label{eq:scaling2}
	\end{equation}
where $\mu$ is the viscosity

\subsection{Geometric models}
Three-dimensional models of the channels were created using OnShape (Cambridge, MA), a computer-aided design (CAD) software package. 
The cross-sections shown in figure \ref{fig:model} were sketched in a 2D plane. The dimensions for each model were prescribed using the dimensionless
lengths earlier stated. Each sketch was then extruded such that the annulus length included two full wavelengths of the peristaltic wave. 

Five models were created for the elliptical annulus simulations with $\alpha/\beta$ equal to $1.0$ (concentric circular), $1.067$, $1.224$, $1.420$, and $1.667$.
For the eccentric simulations, the concentric circular annulus model from above was used as the base model. 
The outer boundary was then offset from the center of the inner circle. Five eccentric circular models were made by setting $\epsilon$ to 0.075, 0.150, 0.225, 0.300, or 0.349. 
The most eccentric model was chosen to have the the same minimum distance between the inner and outer boundaries as in the most elliptic model.

\subsection{Simulations}

\begin{figure}
	\centerline{\includegraphics[height=3.5cm]{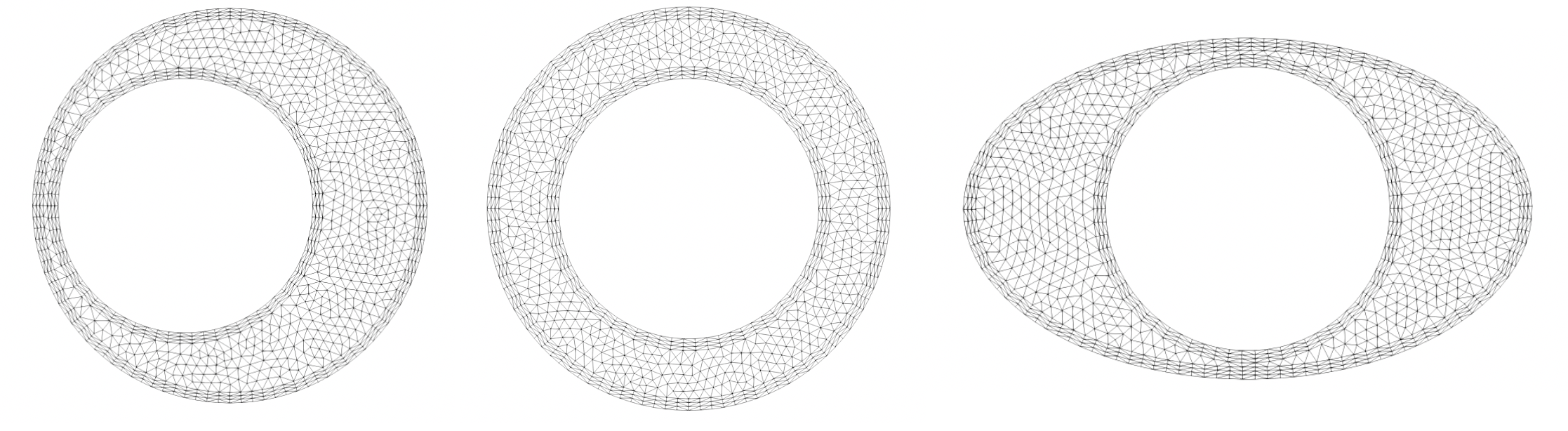}}
	\caption{The `Finer, Fine BL' meshing scheme (bold in table \ref{tab:mesh_convg}) applied to the cross-sections of the most eccentric circular annulus (left), 
	the concentric circular annulus (center), and the most elliptical concentric annulus (right).}
	\label{fig:mesh_schema}
\end{figure}

Models were  exported in the PARASOLID format from OnShape and imported into SimVascular (http://www.simvascular.org; \cite{updegrove2017}) and meshed with MeshSim (Simmetrix, Inc., Troy, NY). 
Regions near the inner and outer walls of the perivascular space were  refined with a boundary layer (BL) mesh. 
Mesh parameters (global edge size, BL height, BL edge size) were determined with a convergence study, results of which are shown in table \ref{tab:mesh_convg}. 
The convergence study was performed using the eccentric model $(\epsilon = 0.349)$. Each simulation was run with a wave speed of $16 {\rm \mu m/s}$, frequency $1 {\rm Hz}$, and an amplitude of $1\%$ of the inner diameter.
The global edge size, BL height, and BL size of the `Finer, Fine BL' simulation was selected because mean flow was within $1\%$ of the finest model and computational time was $10$ hours faster for a single period than the `Finest' simulation. 
A similar mesh scheme was  applied to the remaining elliptical and eccentric models (figure \ref{fig:mesh_schema}).

\begin{table}
    \centering
    \resizebox{\textwidth}{!}{\begin{tabular}{c|c|c|c|c|c}
         \bf{Simulation Name} & \bf{Number of Elements}  {\boldmath ($N_{Elems}$)}& \bf{Global Edge Size} {\boldmath ($L^*_{global}$)} & \bf{BL Height}  {\boldmath ($H^*_{BL}$)} & \bf{BL Edge Size}  {\boldmath ($L^*_{BL}$)} & \bf{net {\boldmath $Q^*$}}  \\
         Coarse, No BL & $510,000$ & $0.15$ & $0$ & $0$ & $1.89e-2$ \\
         Coarse, Coarse BL & $760,00$ & $0.15$ & $0.05$ & $0.05$ & $1.96e-2$ \\
         Coarse, Fine BL & $1,200,000$ & $0.15$ & $0.075$ & $0.025$ & $1.99e-2$ \\
         Medium, No BL & $1,360,00$ & $0.10$ & $0$ & $0$ & $1.98e-2$ \\
         Medium, Coarse BL & $2,000,000$ & $0.10$ & $0.05$ & $0.05$ & $2.02e-2$ \\
         Medium, Fine BL & $2,300,000$ & $0.10$ & $0.1$ & $0.05$ & $2.03e-2$ \\
         Averaged, Coarse BL & $1,200,000$ & $0.12$ & $0.05$ & $0.05$ & $2.00e-2$ \\
         Averaged, medium BL & $2,000,000$ & $0.12$ & $0.075$ & $0.025$ & $2.02e-2$ \\
         \bf{Finer, Fine BL} & $2,100,000$ & $0.10$ & $0.10$ & $0.03$ & $2.03e-2$ \\
         Finest, No BL & $3,000,000$ & $0.07$ & $0$ & $0$ & $2.04e-2$\\ 
    \end{tabular}}
    \caption{Convergence study to determine the number of elements required to accurately simulate the flow
    in the model with the narrowest gap width, the eccentric circular annulus model with eccentricity $\epsilon = 0.349$.
    Flow was driven by a wall wave with wave speed $16 { \rm \mu m/s}$, frequency $1 {\rm Hz}$, wavelength $16 { \rm \mu m}$, with a model length equal to two full wavelengths $(32 { \rm \mu m})$.
    All lengths have been nondimensionalized using the inner radius, e.g. $L_{edge}^* = L_{edge}/r_1$.}
    \label{tab:mesh_convg}
\end{table}

All simulations were performed using Multi-Physics Finite Element Solver (MUPFES, \cite{esmaily2013, esmaily2015}) assuming non-porous walls and an incompressible Newtonian fluid using the average properties for water at $37^{\circ}\rm{C}$ (dynamic viscosity $0.693$ ${\rm mPa\cdot s}$ and density $994$ ${\rm kg / m^3}$). 
On average, arterial displacements are measured to be $1\%$ of arterial diameter \textit{in vivo} \citep{mestre2018}.
We replicate this in our simulations by prescribing an inner wall wave of amplitude $b^* = 0.02$.
In our simulations we assume the frequency to be that of a typical mouse heartbeat$(5$ ${\rm Hz})$, a wave speed of $1000$ ${\rm \mu m /s}$, and wavelength of $200$ ${\rm \mu m} $. 
A traction-free (Neumann) boundary condition was prescribed at  the inlet and outlet, such that there was no imposed pressure gradient and flow was soley driven by the peristaltic wave. 
The local hydrodynamics were then solved by MUPFES using the fully coupled time-dependent Navier-Stokes equation. 
Mean flows converged on the third cycle of the peristaltic wave, but all simulations were run for five cycles to ensure cyclic convergence was reached. 
All simulations were temporally converged with 200 time steps per period of the peristaltic wave.

\subsubsection{Validation for steady flow}

\begin{figure}
	\centerline{\includegraphics[height=4.25cm]{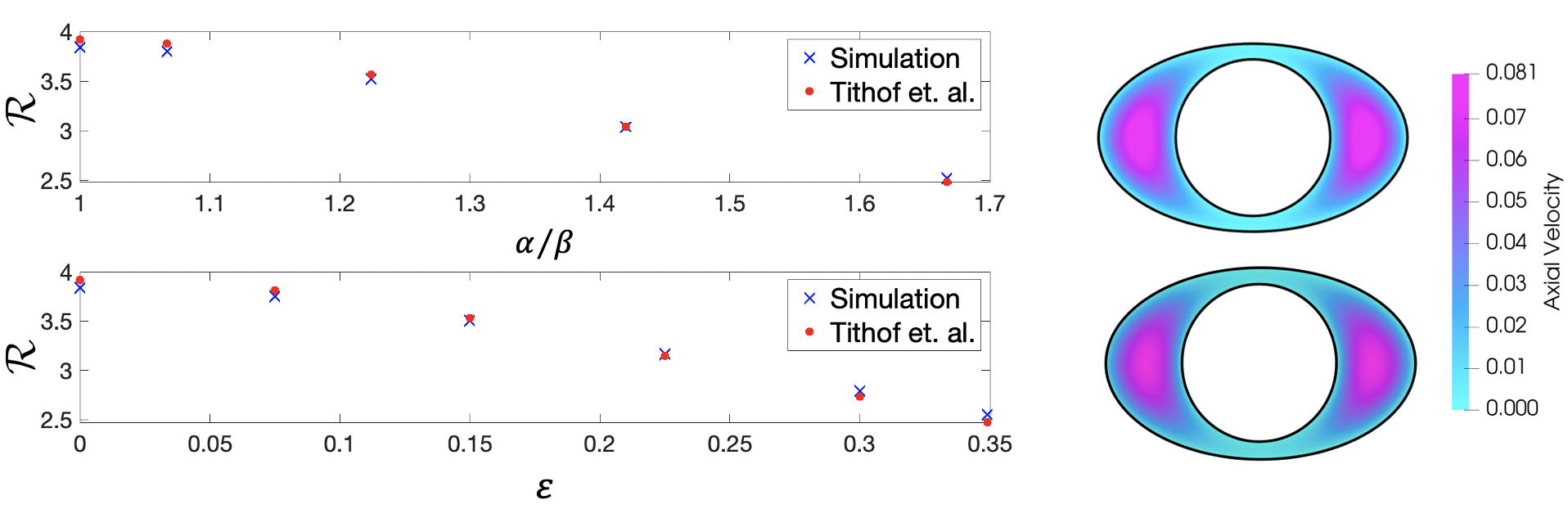}}
	\caption{ Hydraulic resistances (dimensionless) for steady flow calculated using our 3D code and compared with 
	the corresponding values calculated by \citet{Tithof2019}. (left) Upper plot: hydraulic resistances of concentric
	elliptical annuli of different ellipticity $\alpha/\beta$. Lower plot: hydraulic resistances of eccentric circular annuli 
	of different eccentricity $\epsilon$
	(right) Axial velocity profile comparison between \citet{Tithof2019} (upper) and a slice from our model (lower) plotted using the same same scale bar.
	Maximum axial velocity agrees within $1\%$.
	}
	
	\label{fig:resistance}
\end{figure}

As one test of our 3D code, we simulated steady, pressure-driven flow (Poiseuille flow), without a wall wave, in several models and compared the results with the 2D calculations of \citet{Tithof2019} for infinitely long channels with the same cross-sections.
For each 3D simulation we calculated the the hydraulic resistance $\mathcal{R} = (- {dp/dz)/{Q}}$,  
where $dp/dz$ is the axial pressure gradient and $Q$ is volume flow rate, and compared the values to the corresponding values given by \citet{Tithof2019}. Figure \ref{fig:resistance} shows the results of these comparisons. All the hydraulic resistance values calculated with our 3D code agree with the values of \citet{Tithof2019} to within $3\%$. 
Additionally, axial velocities were compared between the two models and agreed within $1\%$ for all models.

\subsubsection{Validation for unsteady, peristaltic flow}

\begin{figure}
	\centerline{\includegraphics[height=5cm]{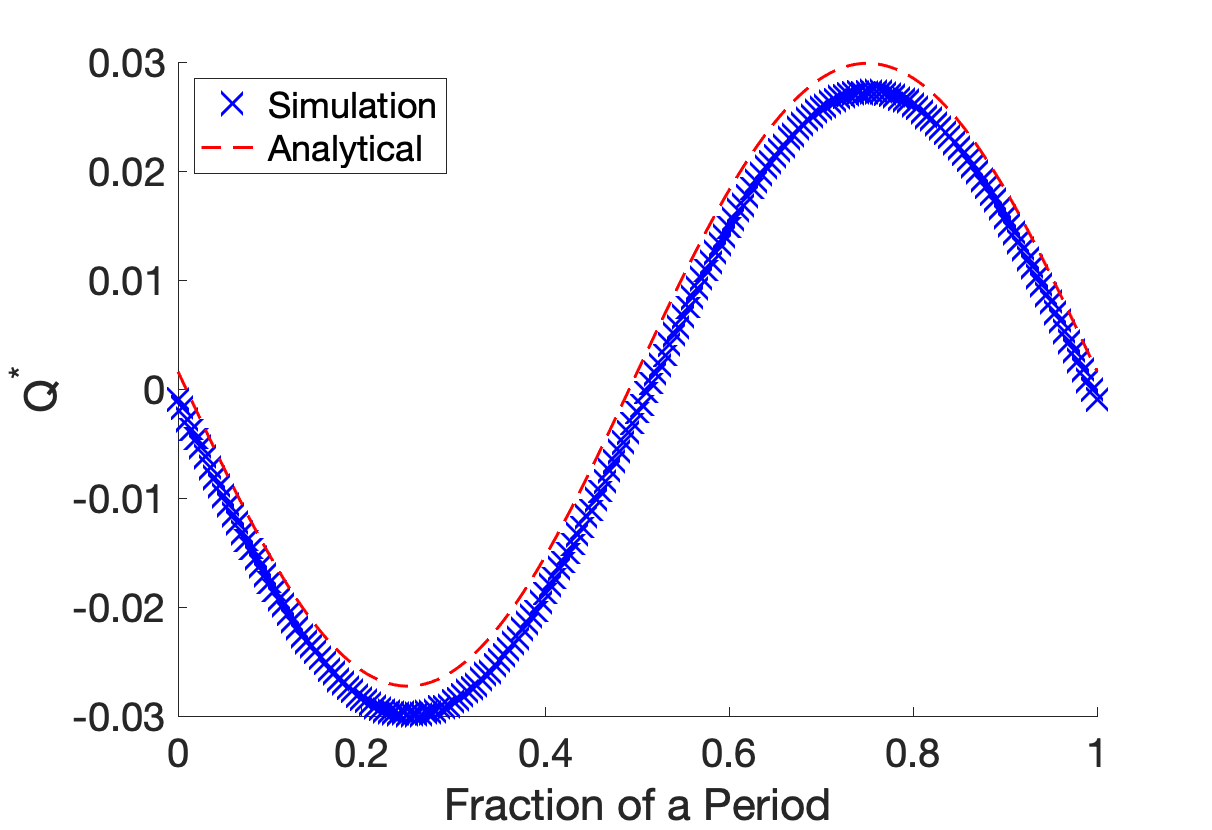}}
	\caption{ Comparison of the instantaneous volumetric flow rate $(Q^*)$ of the simulation with that of the analytical solution, 
	plotted for one period of the peristaltic wave. 
	A peristaltic wave of wave speed $1000$ ${\rm \mu m/s}$, frequency $ 5$ ${\rm Hz}$, wavelength $200$ ${\rm \mu m}$, and amplitude $b^* = 0.02$ was propagated along the inner boundary of the concentric circular annulus model with area ratio of $1.4$, $\alpha = \sqrt{2.4}$, and a model length equal to two wavelengths of the traveling wave.}
	\label{fig:jia_inst}
\end{figure}

In order to test our 3D code in the case of an unsteady flow, we simulated peristaltic pumping in a concentric circular annulus and compared
the results with the analytical solution for an open concentric circular annulus presented here in appendix \ref{appA}. A peristaltic wave of wave speed $1000$ ${\rm \mu m/s}$, frequency $ 5$ ${\rm Hz}$, wavelength $200$ ${\rm \mu m}$, and amplitude $b^* = 0.02$ was propagated along the inner boundary of the concentric circular annulus model with area ratio of $1.4$, $\alpha = \sqrt{2.4}$, and a model length equal to two wavelengths of the traveling wave.
The instantaneous volumetric flow rates $(Q^*)$ for the simulation and the corresponding analytical solution are plotted in figure \ref{fig:jia_inst}. 
The time-averaged volumetric flow rates agree to within approximately $1.25\%$.

\section{Results}

\subsection{The concentric elliptical annulus}

\begin{figure}
	\centerline{\includegraphics[height=8cm]{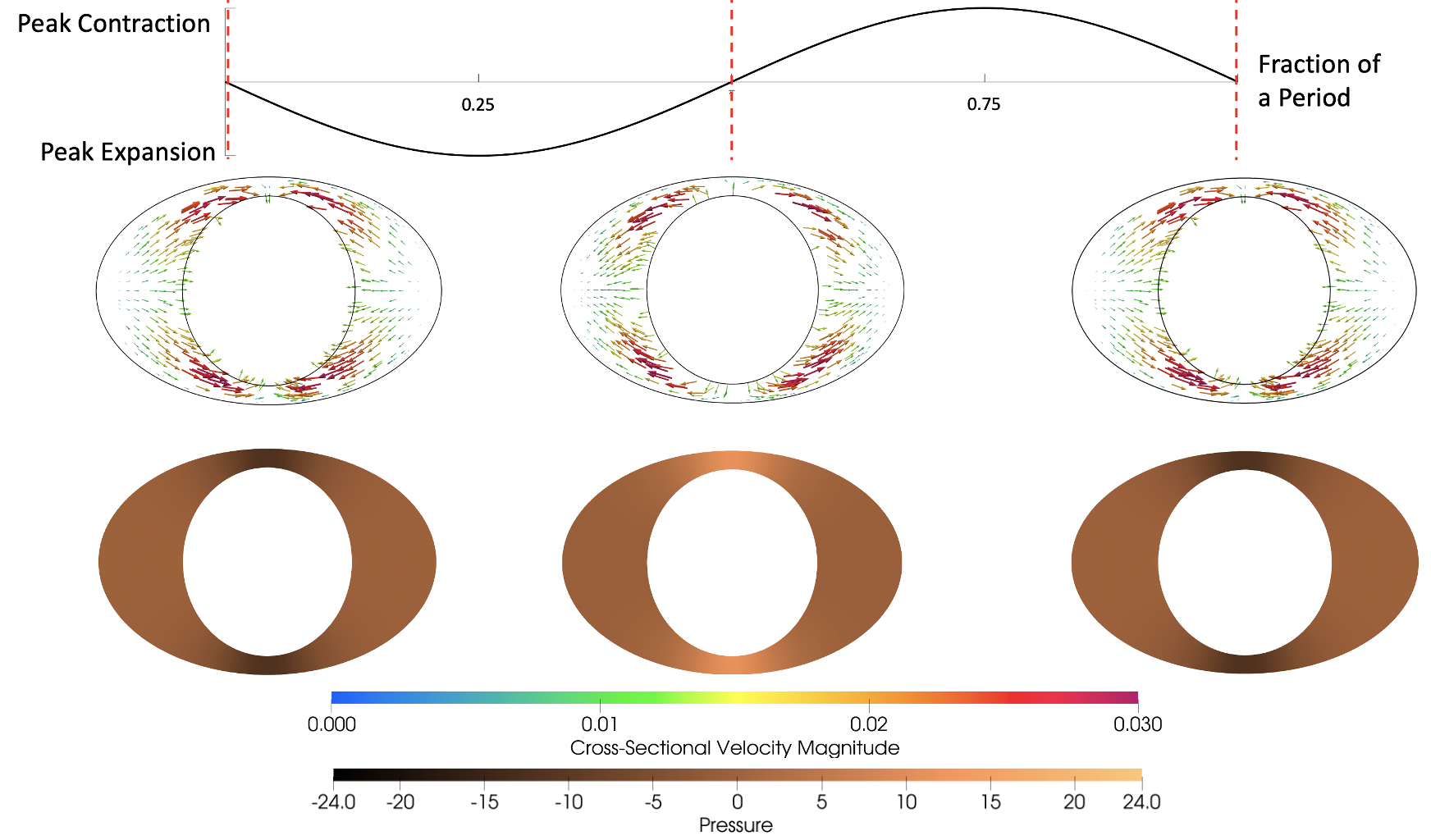}}
	\caption{Cross-sectional pressure distribution and velocity fields at the middle cross-section (one wavelength from either end) of the
	concentric elliptical annulus model with ellipticity ${\alpha}/{\beta} = 1.667$. Values are plotted at three different phases of the peristaltic 
	wave: beginning, middle, and end (left to right). The displacement of the wall wave is plotted above at the top, with red vertical dashed lines 
	marking the three times of the corresponding cross-sections. The color scale indicates the magnitude of the dimensionless velocity magnitude,
	and the copper scale indicates the dimensionless pressure.}
	\label{fig:velocity-pressure}
\end{figure}

First we examine the results for concentric annuli with an elliptical outer boundary. 
Introducing this ellipticity induces an azimuthal pressure gradient in each axial cross-section along the length of the tube, 
which drives an azimuthal velocity component. 
In the elliptical models, the flow oscillates in a `four lobe' pattern in the quadrants created by the major and minor axes of the ellipse. 
The narrowest regions of each axial cross-section are at the ends of minor axis of the ellipse: henceforth we refer to these as the `narrow gaps.' 
The widest regions of each cross-section occur at the ends of the major axis of the ellipse: we call these the `wide gaps.'
In figure \ref{fig:velocity-pressure}, we show the cross-sectional velocity and pressure distribution at three times during a cycle of the peristaltic wave: 
beginning, middle, and end of a single wave period. These times are marked with the red vertical dashed lines along the plot of the peristaltic wave (top). 
In order to plot the cross-sectional pressure distribution, we subtract the mean pressure over each slice in order to remove the axial pressure variation. 
The resulting pressure is then nondimensionalized as

\begin{equation}
	p_{dist}^* =(p - p_{mean}) \frac{\lambda}{\mu c} .
	\label{eq:pres_dist}
\end{equation}

At the beginning of the cycle, the inner boundary is contracting from its mean position. 
This results in low pressure in the narrow gaps and high pressure in the wide gaps, thus driving fluid toward the narrow gaps. 
When the inner boundary reaches its peak contraction, the azimuthal pressure gradient and velocity both go to zero. 
Then, as the inner boundary expands, high pressure develops in the narrow gaps and low pressure in the wide gaps, 
driving fluid out of the narrow gap and into the wide gap.When the inner boundary reaches its peak expansion, the 
azimuthal pressure gradient and velocity again go to zero. This cycle then repeats during each period of the wall wave. 

\begin{figure}
	\centerline{\includegraphics[height=8cm]{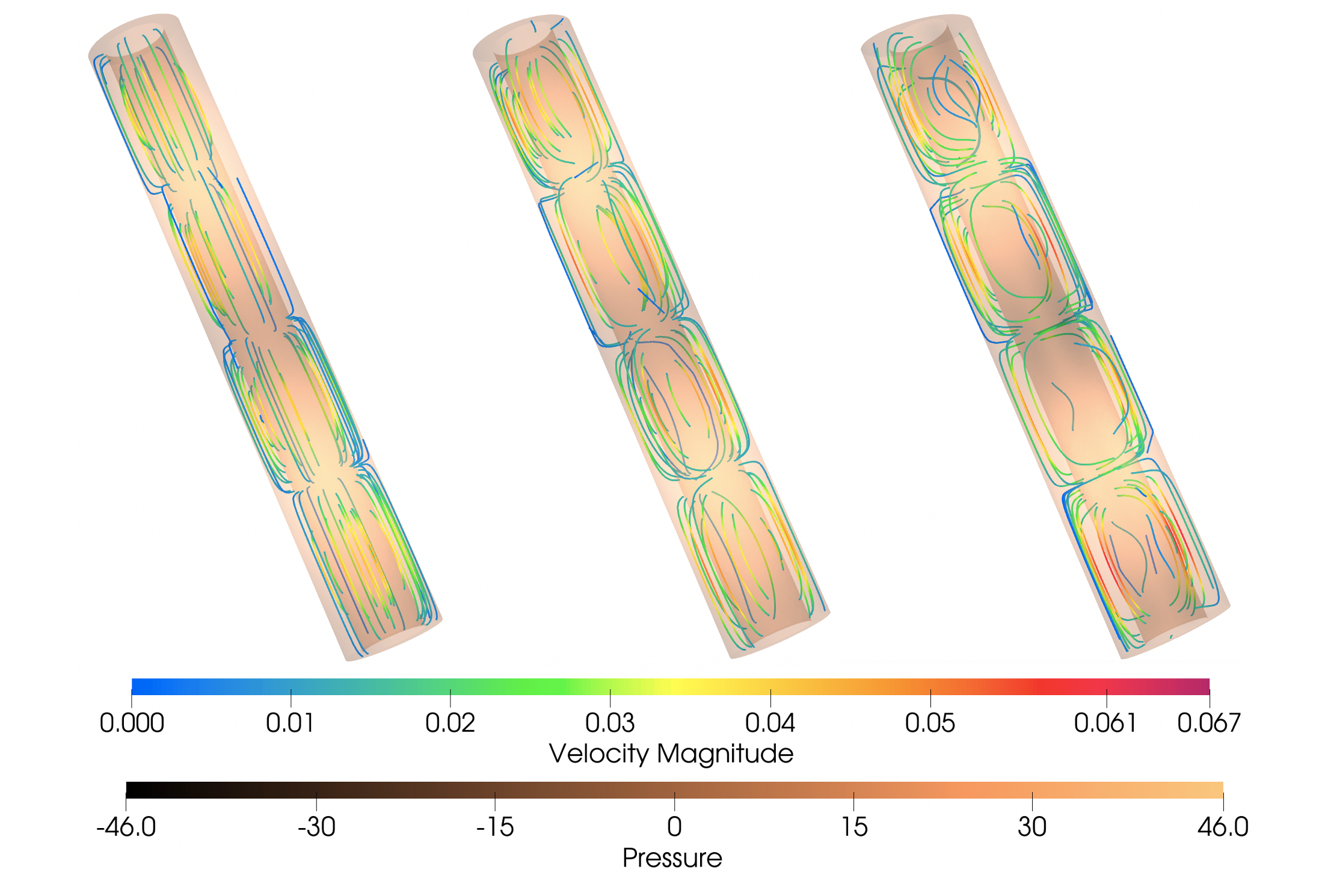}}
	\caption{Instantaneous streamlines for pumping in concentric annuli with ellipticity ${\alpha}/{\beta}$ equal to $1.00$ (circular), $1.224$, and $1.667$. 
	For the concentric circular annulus (left) the flow is axisymmetric and the streamlines wiggle only in the radial direction. When the outer 
	wall of the annulus is slightly flattened into an ellipse (center), the flow becomes three-dimensional, with an oscillating azimuthal velocity component, 
	and the streamlines also wiggle in the azimuthal direction. For substantial flattening (right), the azimuthal velocity is significant, as are the azimuthal 
	wiggles in the streamlines. The color and copper scales indicate values of the dimensionless velocity magnitude and pressure.}
	\label{fig:ellip_streamlines}
\end{figure}

The resulting azimuthal flow causes the velocity field to be fully three-dimensional, with components in the radial, azimuthal and 
axial directions, and the instantaneous streamlines wiggle in the azimuthal as well as the radial direction. 
Figure \ref{fig:ellip_streamlines} shows instantaneous streamlines for a concentric circular annulus ($\alpha/\beta = 1$)
and two of the elliptical annuli ($\alpha/\beta = 1.224$ and $1.667$). 
In the concentric circular annulus (left in figure \ref{fig:ellip_streamlines}), with no azimuthal velocity,  the streamlines are axisymmetric, lie in planes through the central axis, and 
wiggle only in the radial direction.  As the outer boundary becomes elliptical, the streamlines develop an azimuthal wiggle due to the azimuthal velocity component.
In the moderately elliptical annulus (central in figure \ref{fig:ellip_streamlines}), a slight azimuthal wiggle can be seen along each streamline. 
In the more elliptic model (right in figure \ref{fig:ellip_streamlines}), the streamlines are show significant azimuthal wiggles due to
a stronger azimuthal velocity component.
In table \ref{tab:ellip_rad_azi_vel}, we quantify the radial $(v_r)$ and azimuthal $(v_{\theta})$ velocities and the ratio between the two for all of the elliptical annuli simulations. As the streamline plots suggest, the azimuthal velocity increases with increasing ellipticity and becomes approximately equal in magnitude 
to the radial component at $\alpha /\beta = 1.420$.

\begin{table}

    \centering
    \scalebox{0.5}{
    \resizebox{12cm}{!}{\begin{tabular}{c|c|c|c}
         \bf{$\alpha / \beta$} & {$\overline{v_{r}}_{\rm max}$}  & {$\overline{v_{\theta}}_{\rm max}$} & 
         {$\overline{v_{\theta}}_{\rm max} / \overline{v_{r}}_{\rm max}$} \\
         $1.000$ & $5.00e-3$ & $0.00$ & $0.00$ \\
         $1.067$ & $4.93e-3$ & $1.39e-3$ & $0.28$ \\
         $1.224$ & $5.54e-3$ & $4.18e-3$ & $0.78$ \\
         $1.420$ & $6.09e-3$ & $6.75e-3$ & $1.11$ \\
         $1.667$ & $7.03e-3$ & $8.95e-3$ & $1.27$ \\
         
    \end{tabular}}}
        \caption{Comparison of the magnitudes of the radial $(v_r)$ and azimuthal $(v_{\theta})$ velocities (dimensionless) for each of 
        the elliptical models. Each component was averaged over a cross-section for each point in time during the peristaltic wave 
    ($\overline{v_{r}}$, $\overline{v_{\theta}}$), and the maximum values of these averages during the period 
    ($\overline{v_{r}}_{\rm max}$, $\overline{v_{\theta}}_{\rm max}$), and their ratio, are  
    listed in the table.}
    \label{tab:ellip_rad_azi_vel}
\end{table}

\subsection{The eccentric circular annulus}

\begin{figure}
	\centerline{\includegraphics[height=8cm]{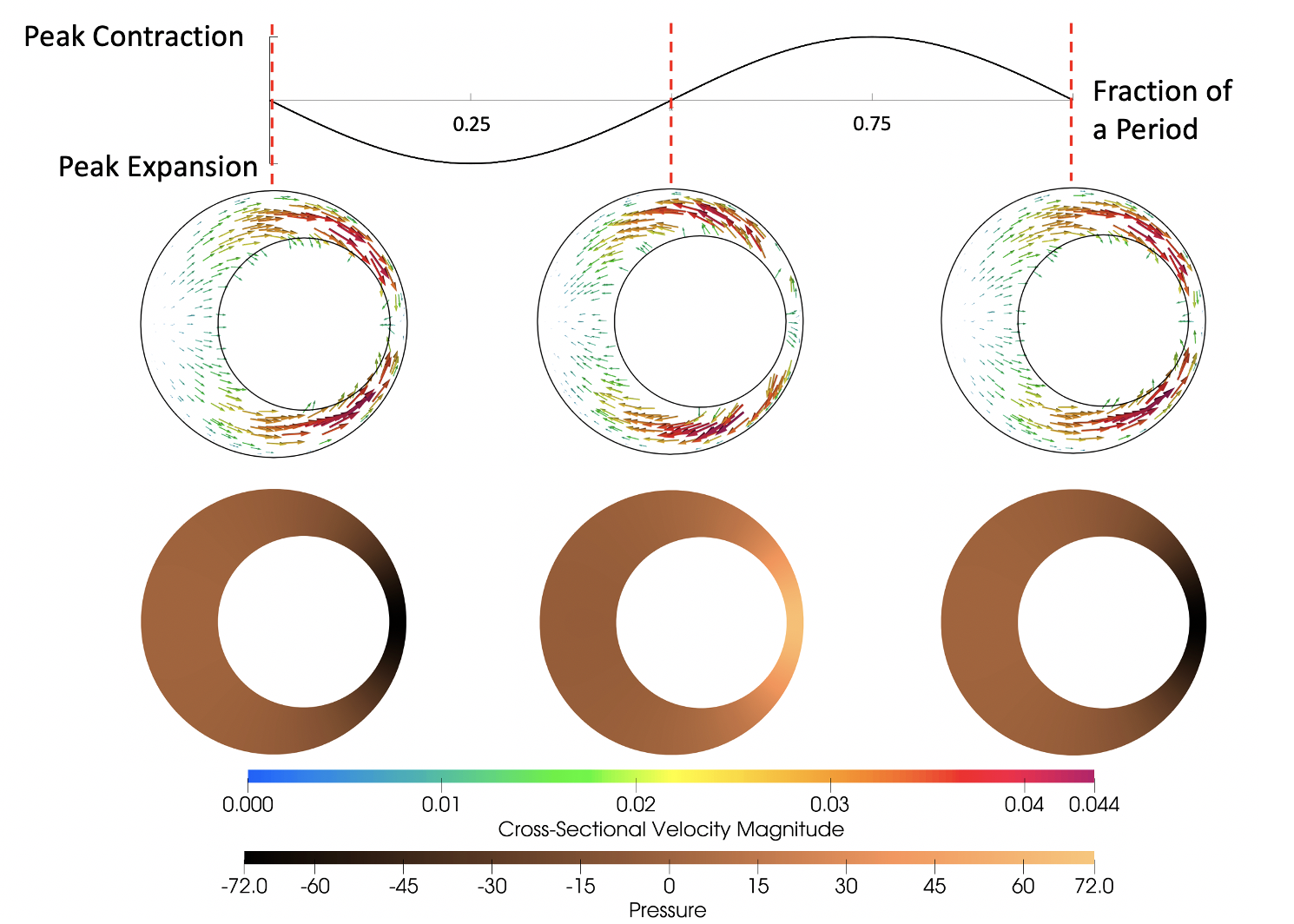}}
	\caption{As in figure \ref{fig:velocity-pressure}, but for an eccentric circular annulus model with eccentricity $\epsilon = 0.349$.} 	
	\label{fig:ecc_velocity-pressure}
\end{figure}

Next we turn to the results for the circular annuli with increasing eccentricity. 
Figure \ref{fig:ecc_velocity-pressure} shows cross-sectional pressure distributions (equation \ref{eq:pres_dist}) and velocity when the inner boundary is offset.
The flow oscillates in a `two lobe' shape, separated by a line of symmetry in the same direction as the eccentricity of the inner boundary. 
By offsetting the inner boundary, a narrow gap and a wide gap are formed at the ends of this line of symmetry. 
As in the elliptical annulus, the pressure is initially low in the narrow gap as the inner boundary contracts from its mean position.
When the inner boundary reaches peak contraction, the azimuthal pressure gradient and velocity go to zero.
The inner boundary then begins to expand, causing higher pressure to form in the narrow gap and lower pressure to form in the wide gap, driving an azimuthal velocity
toward the wide gap. Once the inner boundary reaches peak expansion, the azimuthal pressure gradient and flow go to zero. and then reverse as the 
inner boundary begins to contract again. This cycle then repeats for each period of the traveling wall wave. 

\begin{figure}
	\centerline{\includegraphics[height=8cm]{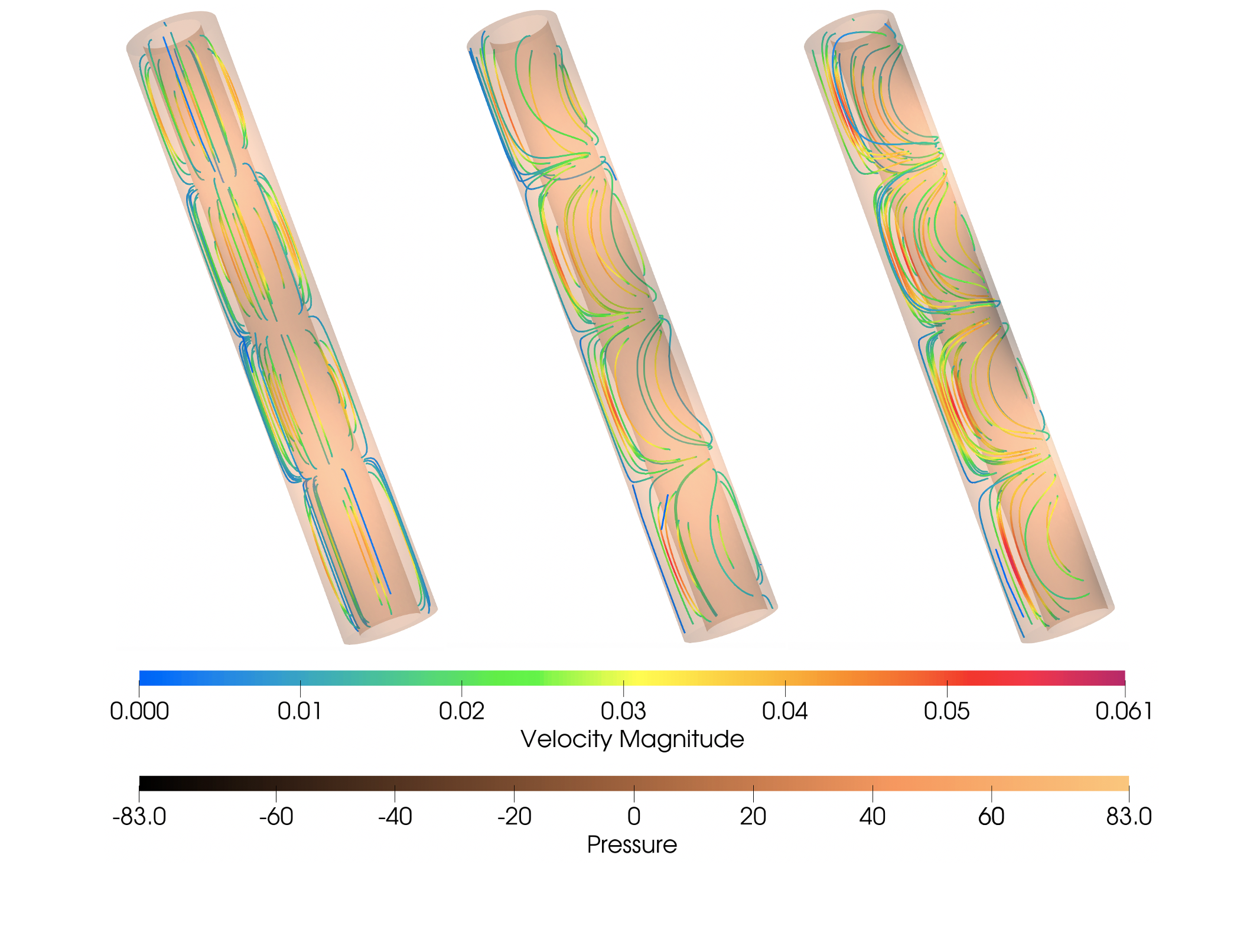}}
	\caption{As in figure \ref{fig:ellip_streamlines}, but for circular annuli with eccentricity $\epsilon$ equal to $0$, $0.225$, and $0.349$. The instantaneous
	streamlines wiggle only in the radial direction in the concentric annulus (left), but also wiggle increasingly in the azimuthal direction with increasing 
	eccentricity (center, right).}
	\label{fig:ecc_streamlines}
\end{figure}

As in the elliptical case, the instantaneous streamlines wiggle in both the radial and azimuthal directions. 
Figure \ref{fig:ecc_streamlines} shows instantaneous streamlines for the concentric circular annulus and two eccentric cases $(\epsilon = 0.225$ and $0.349)$
 plotted at the same phase of the peristaltic wall wave. When eccentricity is introduced, the streamlines begin to wiggle in the azimuthal direction, 
as an azimuthal motion is driven by the pressure distribution, as illustrated in figure \ref{fig:ecc_velocity-pressure}. 
As the eccentricity is increased, the streamlines wiggle more in the azimuthal direction and the ratio of azimuthal to radial velocities increases. 
The radial and azimuthal velocities are approximately equal in magnitude at ellipticity $\epsilon = 0.150$ (table \ref{tab:ecc_rad_azi_vel}).

\begin{table}
    \centering

    \quad \quad
    \scalebox{0.5}{
    \resizebox{12cm}{!}{\begin{tabular}{c|c|c|c}
         \bf{$\epsilon$} & {$\overline{v_{r}}_{\rm max}$}  & {$\overline{v_{\theta}}_{\rm max}$} & 
         {$\overline{v_{\theta}}_{\rm max} / \overline{v_{r}}_{\rm max}$} \\
         $0.000$ & $5.70e-3$ & $0.00$ & $0.00$ \\
         $0.075$ & $5.84e-3$ & $2.54e-3$ & $0.43$ \\
         $0.150$ & $6.01e-3$ & $5.00e-3$ & $0.83$ \\
         $0.225$ & $6.28e-3$ & $7.70e-3$ & $1.23$ \\
         $0.300$ & $6.61e-3$ & $1.02e-2$ & $1.54$ \\
         $0.349$ & $6.93e-3$ & $1.22e-2$ & $1.76$ \\
         
    \end{tabular}}}
    \caption{Comparison of the magnitudes of the radial $(v_r)$ and azimuthal $(v_{\theta})$ velocities (dimensionless), as in table \ref {tab:ellip_rad_azi_vel}, 
    but for the eccentric circular annulus models of different eccentricity $\epsilon$.}
    \label{tab:ecc_rad_azi_vel}
\end{table}

\subsection{The mean volumetric flow rate}

\begin{figure}
	\centerline{\includegraphics[height=8cm]{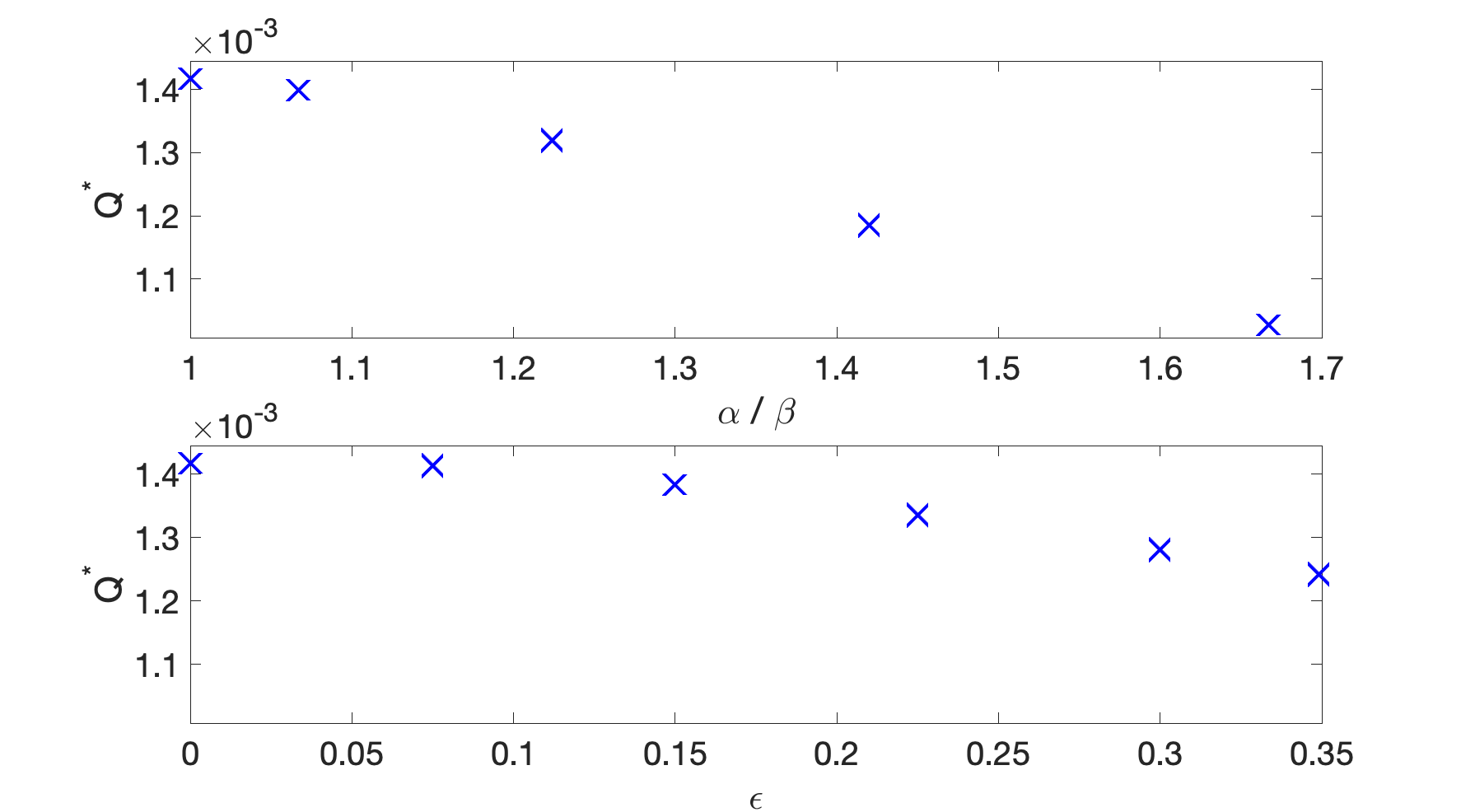}}
	\caption{ Time-averaged volumetric flow rate (dimensionless) plotted as a function of the ellipticity of the elliptic annulus model (top),
	and as a function of eccentricity of the eccentric circular annulus model (bottom).}
	\label{fig:vol_flow_ecc}
\end{figure}

The mean (time-averaged) volumetric flow rate $Q^*$ was calculated for each of the elliptical and eccentric models (figure \ref{fig:vol_flow_ecc})
by averaging over a full wave period. In all
cases, $Q^*$ is in the direction of propagation of the wall wave. For the elliptic models, $Q^*$ decreases monotonically with increasing ellipticity. For
the eccentric circular models, $Q^*$ decreases monotonically with increasing eccentricity.

For a concentric circular annulus, the pumping effectiveness scales as $(b/\ell)^2$. When either ellipticity or eccentricity is introduced, 
the gap width $\ell$ varies azimuthally, and this scaling suggests that the pumping effectiveness also varies azimuthally, being greater where the 
gap width $\ell$ is smaller. Based solely on this scaling, it is not immediately obvious whether the net effect of introducing a small amount ellipticity
or eccentricity is to increase or decrease the total pumping effectiveness. The results of our simulations show that, for fixed cross-sectional area
of the annulus, the pumping effectiveness always decreases when either ellipticity of eccentricity is introduced, and hence the concentric circular
annulus provides the most efficient configuration for pumping. 

We know that for steady, pressure-driven flow in an annular tube of fixed cross-sectional area, the hydraulic resistance decreases with increasing 
ellipticity or eccentricity. \citet{Tithof2019} speculated that this decrease in viscous resistance might offset to some extent the accompanying decrease 
in peristaltic pumping efficiency, such that optimal pumping occurs in annular cross-sections that are not concentric circles. Here we have shown that 
this is not the case: apparently the decrease in pumping efficiency dominates the reduction in viscous resistance when any amount of ellipticity 
or eccentricity is introduced. 

\section{Discussion}

% In this study, we have simulated flows that are driven only by the motion of the inner wall of the annulus; the tube is open to ambient pressure at both ends. 
% Similar zero pressure-drop boundary conditions have been used in some simulations \citep{asgari2016, kedarasetti2020}, while others have also imposed a small 
% constant pressure gradient applied along the length of the tube \citep{bilston2003,  kedarasetti2020}.
% In a realistic physiological system, however, such simple pressure conditions do not exist: the peristaltic pump is embedded in a larger system that has 
% additional flow components upstream and downstream.
% Meaningful comparisons of models with experimentally observed flows would require time-dependent boundary conditions that reflect the presence of resistance 
% and compliance elsewhere in the system; arterial (and other tissue) compliance acts as an elastic-energy reservoir that reduces the oscillatory amplitude 
% of flow downstream of the model tube. Coupling realistic boundary conditions, including resistance and compliance, to previous simulations and models of 
% perivascular pumping yields flows whose oscillatory dynamics agree well with {\it in vivo} flow measurements \citep{ladron2020, mestre2018}.

In this study, we have simulated flows that are driven only by the motion of the inner wall of the annulus; the tube is open to ambient pressure at both ends.
Similar zero pressure-drop boundary conditions have been used in some simulations, while others have also imposed a small constant pressure gradient applied along the length of the tube \citep{asgari2016, bilston2003,  kedarasetti2020}. 
The resulting flow has a large amplitude compared to the mean (fig. \ref{fig:jia_inst}), which is not observed {\it in vivo} \citep{mestre2018} and has led some to conclude that peristalsis does not drive perivascular flow \citep{kedarasetti2020}. 
In a realistic physiological system, however, such simple pressure conditions do not exist: the peristaltic pump is embedded in a larger system that has 
additional flow components upstream and downstream.
Meaningful comparisons of models with experimentally observed flows would require time-dependent boundary conditions that reflect the presence of resistance and compliance elsewhere in the system;
arterial (and other tissue) compliance acts as an elastic-energy reservoir that reduces the oscillatory amplitude of flow downstream of the model tube and shields the peripheral circulation from high oscillatory shear stress.
Coupling realistic boundary conditions, including resistance and compliance, to previous simulations and models of perivascular pumping yields flows whose oscillatory dynamics agree well with {\it in vivo} flow measurements \citep{ladron2020, mestre2018}.

Here, with our simple open-end boundary conditions, we have focused on the effect of non-axisymmetry of the cross-section in producing three-dimensional 
peristaltic flow in an annular tube. 
Asymmetry of the cross-section causes azimuthal pressure gradients which drive an oscillatory azimuthal flow that is not present in the concentric circular annulus.
This is a direct result of the azimuthal variation of the gap width, and hence the pumping effectiveness, around the annulus. 
In the widest portions of the annulus, hydraulic resistance is at its lowest, but pumping effectiveness is also at its lowest. 
In the narrow regions of the annulus, flow resistance is higher, but the pumping is more effective. The net effect, however, is that, for fixed
area of the annulus cross-section, overall pumping effectiveness, as measured by the mean volumetric flow rate,  is reduced when either ellipticity or eccentricity are introduced.
%When the wall wave is narrowing an axial cross-section (middle in figures \ref{fig:ecc_velocity-pressure}, \ref{fig:velocity-pressure}) high pressure is formed 
%in the narrow gap of the model and fluid is pushed towards the wider portions because there is lower resistance further from the walls.
%When the displacement wave is at its peak, both flow and cross-sectional pressure distributions are zero. 
%As the wave begins to widen the cross-section, a region of low pressure is formed in the narrow gaps of the model causing the flow to reverse and flow from the widest portions to the %narrowest (final in figures \ref{fig:ecc_velocity-pressure}, \ref{fig:velocity-pressure}). 
%Globally, this causes streamlines to wiggle in the azimuthal direction (figures \ref{fig:ellip_streamlines}, \ref{fig:ecc_streamlines}) in the four-lobe or two-lobe shape for elliptical and %eccentric annuli, respectively. 

One of the interesting effects of the three-dimensional nature of the flows we consider here is their contribution to
Taylor dispersion \citep{taylor1953, aris1956} of a solute. \citet{asgari2016} have shown how a purely oscillatory shear flow 
(with no net mean flow) enhances dispersion in the axial direction, compared to pure diffusion in the absence of a flow. (This enhancement is weak, however, 
compared to the classical Taylor dispersion due any net axial flow induced by the pumping: see the discussion of this point in \citet{thomas2019}.) 
Here we point out that, for peristaltic flows in non-axisymmetric annuli, the oscillating shear flow in the azimuthal direction will enhance dispersion
in that direction, in cases where the solute concentration is not initially axisymmetric. For example, consider an annulus 
in which a solute is injected through a small port in the outer wall. The solute enters locally at one particular azimuthal position,
and if the annulus is axisymmetric, the solute will spread azimuthally only by ordinary diffusion. If, however, the annulus
is non-axisymmetric, then the oscillating shear flow in the azimuthal direction will enhance the dispersion in that direction,
establishing more quickly a uniform solute distribution around the annulus. This enhancement of mixing might have 
useful industrial or biomedical applications.

\bigskip
\section*{Acknowledgements}

We thank our colleagues Douglas Kelley and Jeffrey Tithof for many helpful discussions of this research. This work was supported by the NIH/National Institute of Aging (grant RF1AG057575) and by the U. S. Army Research Office 
(grant MURI W911NF1910280). The views and conclusions contained in this article are solely those of the authors and should not be interpreted 
as representing the official policies, either expressed or implied, of the National Institutes of Health, the Army Research Office, 
or the U.S. Government. The U.S. Government is authorized to reproduce and distribute reprints for Government purposes 
notwithstanding any copyright notation herein.

\appendix
\section{Analytical solution for a concentric circular annulus}
\label{appA}

We present here an analytical solution of the problem of peristaltic pumping in a thin, open tube with cross-section in the
form of a concentric circular annulus. \citet{wang2011} considered the related problem of peristaltic pumping a concentric circular 
annulus filled with a porous medium, with flow driven by a sinusoidal wall wave propagating along the inner boundary. They
use the long wavelength, low Reynolds number approximation \citep{shapiro1969, jaffrin1971} and  are able to obtain an 
analytical solution. Their solution for the time-averaged volume flow rate $Q$ is the following:
%\begin{linenomath*}
	\begin{eqnarray}
	Q = \pi r_2^2  \gamma c \left(  \frac{2 \alpha^2}{1 - \alpha^2} \right)  \left( \frac{b}{r_2} \right)^2  %\nonumber \\
	+ \pi r_2^2 (1-\alpha^2 ) \left( - \frac{\kappa \Delta p_\lambda}{\lambda \mu} \right)  \nonumber \\
	-\pi r_2^2 \left( \frac{1+ 3 \alpha^2}{2(1- \alpha^2 )} \right) \left(-  \frac{\kappa \Delta p_\lambda}{ \lambda \mu} \right) \left( \frac{b}{r_2} \right)^2 ,
	\label{eq:wang}
	\end{eqnarray}
%\end{linenomath*}
where $\alpha = r_1/r_2$ is the ratio of the inner and outer radii of the circular annulus, $\mu$ is the dynamic viscosity, $\gamma$ 
is the porosity of the PVS, $\kappa$ is the Darcy permeability, $\Delta p_\lambda$ is the pressure drop over one wavelength 
$\lambda$, $b$ is the amplitude of the wall wave, and $c$ is the propagation speed of the wall wave. The net volume flow rate 
$Q$ is hence the sum of the three terms on the right-hand side of this equation. 
The first term represents the net flow due to peristaltic pumping by a small-amplitude wall wave in the absence of an overall pressure 
gradient ($\Delta p_\lambda = 0$). The second term represents the net flow driven by an overall pressure gradient in the undisturbed
channel (with no wall wave). The third term represents a correction accounting for the fact that the pressure-driven flow
moves through the annulus whose shape is distorted by the wall wave. (For small wave amplitudes ($b/r_2 << 1$) and a moderate 
imposed pressure gradient, the third term is negligibly small.)
The first term on the right, representing peristaltic pumping, is a purely geometric term
representing the squeezing effect of the wall wave. This term, with $\epsilon = 1$, is the same as one would obtain 
for flow in an open (non-porous) space. This term does not involve the viscosity: the viscosity does 
come into play in determining (based on the second term) what pressure gradient would be necessary to cancel the net peristaltic flow. 

Here we adopt an approach similar to that of \citet{wang2011}, but consider the annular space to be open rather porous medium, 
and hence use the Navier-Stokes equation rather than the Darcy law. We use the same notation as in section 2: for the concentric 
circular annulus, $r_2 = r_3$ and $d_x = 0$. The ratio of the wavelength $\lambda$ to the width of the PVS channel $r_2 - r_1$ is 
assume to be large, the lubrication approximation (very low Reynolds number) is applied. 
As is usual in peristalsis-related problems, a coordinate transformation is adopted: instead of considering a traveling peristaltic wave 
in the laboratory frame, we use the wave frame in which the peristaltic wave is stationary. The transformation between the two frames is
given by
	\begin{equation}
	r^{\prime}=r,\quad z^{\prime}=z-ct,\quad v_r'=v_r,\quad v_z'=v_z-c ,
	\label{eq:wave coor}
	\end{equation}
where quantities in the wave frame are denoted by primes. Considering the long wavelength and the low Reynolds number, 
the governing Navier-Stokes equation in the wave frame is reduced to the following components:
	\begin{equation}
	\label{eq:ns_r}
	\frac{\partial p'}{\partial r'}=0, \\
	\label{eq:ns_z}
	\frac{\partial p'}{\partial z'}=\mu \frac{1}{r}
	\frac{\partial}{\partial r'}(r'\frac{\partial {v_z}'}{\partial r'}).
	\end{equation}
The boundary conditions are:
	\begin{equation}
	\label{eq:bc}
	{v_r}'(h',z')={v_r}'(r_2,z')=0;\quad {v_z}'(h',z')={v_z}'(r_2,z')=-c.
	\end{equation}
From equations \ref{eq:ns_r} and \ref{eq:bc}, the velocity in $z$ direction is obtained:
	\begin{equation}
	\label{eq:vz}
	{v_z}'=\frac{dp'}{dz'}\frac{r'^2}{4\mu}+\frac{\frac{dp'}{dz'}}{4\mu \ln(\frac{h'}{r_2})}[h'^2\ln(\frac{r'}{r_2})-r_2^2\ln(\frac{r'}{h'})]-c.
	\end{equation}
		
The volumetric flow rate $Q'$ is constant, independent of $z'$, which gives
	\begin{equation}
	\label{eq:q0}
	Q'=2\pi\int_{h'}^{r_2}v_z'r'dr'. 
	\end{equation} 
Substituting ${v_z}'$ from equation (\ref{eq:vz}) into equation (\ref{eq:q0}), the pressure gradient can be found in terms of $Q'$. 
Since $h'(z')$ is a periodic function with period $\lambda$, the pressure gradient ${dp'}/{dz'}$ is also periodic
with period $\lambda$, and the pressure drop over a distance $\lambda$,  
	\begin{equation} 
	\label{eq:delta_p0}
	\Delta p'_\lambda=\int_{0}^{\lambda}\frac{dp'}{dz'}dz'.
	\end{equation}
is constant, independent of the starting point of the interval.
Inserting the value of ${dp'}/{dz'}$ given by equations (\ref{eq:vz}) and (\ref{eq:q0}), the relationship between $\Delta p'_\lambda$ and $Q'$ is given as 
	\begin{equation}
	\label{eq:delta_p}
	\Delta p'_\lambda=\int_{0}^{\lambda}-\frac{8\mu \ln(\frac{h'}{r_2})[Q'+\pi c({r_2}^2-{h'}^2)]}{\pi [\ln(\frac{h'}{r_2})({r_2}^4-{h'}^4)+({r_2}^2-{h'}^2)^2]}dz'.
	\end{equation}
Substituting $h'(z')$ with the sinusoidal waveform into equation (\ref{eq:delta_p}), then $\Delta p'_\lambda$ in terms of given $Q'$ can be calculated. Since the integration is too complex to evaluate analytically, it is computed numerically instead. 

Generally, the quantity of practical interest is the time-averaged volumetric flow rate (net flow rate) at each cross-section. In the laboratory frame $\Delta p_\lambda=\Delta p'_\lambda$ and the volumetric flow rate at time $t$ across a cross section at $z$ is 
	\begin{equation}
	\label{eq:Q}
	Q=2\pi\int_{h}^{r_2}v_zrdr=Q'+\pi c({r_2}^2-h^2).
	\end{equation}
The time-averaged flow rate then is given by 
	\begin{equation}
	\label{eq:mean Q}
	\bar Q=\frac{1}{T}\int_{0}^{T}Qdt=Q'+\pi c({r_2}^2-{r_1}^2
	-\frac{1}{2}b^2).
	\end{equation}

\bibliographystyle{jfm}
\bibliography{NewBib.bib}

\end{document}